\documentclass[prb, longbibliography, twocolumn]{revtex4-2}
\usepackage{bm}
\usepackage{graphicx}
\usepackage{amsmath}
\usepackage{amssymb} 
\usepackage[utf8]{inputenc}
\usepackage[T1]{fontenc}
\usepackage{color}
\usepackage{xcolor}
\usepackage{upgreek} 
\usepackage{subfigure}
\usepackage[unicode=true,colorlinks=true,citecolor=blue]{hyperref}
\usepackage{lipsum}
\usepackage{epsfig}
\usepackage{wrapfig}
\usepackage[normalem]{ulem}
\usepackage{units}
\usepackage{cancel}
\usepackage{float}
\usepackage{lipsum}
\usepackage{placeins}
\usepackage{ocgx}
\usepackage{mathtools}

\newcommand{\nix}[1]{}

\newcommand{\pderiv}[2]{\frac{\partial #1}{\partial #2}}
\renewcommand{\phi}{\varphi}

\renewcommand{\i}{\mathrm i}
\newcommand{\e}{\mathrm e}
\newcommand{\eps}{\varepsilon}

\newcommand{\beq}{\begin{equation}}
	\newcommand{\eeq}{\end{equation}}
\newcommand\beqa{\begin{eqnarray}}
	\newcommand\eeqa{\end{eqnarray}}
\newcommand\ba{\begin{array}}
	\newcommand\ea{\end{array}}

\newcommand{\aver}[1]{\left \langle #1 \right \rangle}

\begin{document}
	
\title{
Orbital photoinduced Faraday effect in a lattice of conducting disks}

	\author{A. A. Gunyaga}
	\affiliation{Ioffe Institute, 194021 St. Petersburg, Russia}
	
	\author{M. V. Durnev} 
	\email{durnev@mail.ioffe.ru} 
	\affiliation{Ioffe Institute, 194021 St. Petersburg, Russia}
	
\begin{abstract}

We present a theoretical study of pump-induced Faraday and Kerr rotation in a two-dimensional lattice of conducting disks. We propose a mechanism of Faraday and Kerr responses in which a circularly polarized pump directly induces a high-frequency Hall component $\sigma_{xy}=-\sigma_{yx}$ of the electron conductivity tensor at the probe frequency. Microscopically, $\sigma_{xy}$ originates from the third-order nonlinear response of the electron gas rather than from the real magnetic field created by solenoidal charge currents in the inverse Faraday effect. We show that the Faraday and Kerr rotation angles are significantly enhanced when the pump and probe frequencies are tuned close to the plasmon resonance of the disks, reaching $\sim 0.1^\circ$ per 1~kW/cm$^2$ of incident pump intensity in the terahertz range. This mechanism can explain recently observed giant pump-induced Faraday rotation in graphene disk lattices.

\end{abstract}
	
	\maketitle
	 
	
\section{Introduction}

The inverse Faraday effect (IFE) has attracted considerable interest as a route to light-induced and optically controlled magnetization in nonmagnetic media~\cite{Pitaevskii:1961,Ziel:1965aa,Hareau:2025}. In conducting systems, the IFE magnetization originates from solenoidal charge currents excited by circularly polarized light~\cite{Karpman:1982,Hertel:2006vj}. These currents, and hence the resulting magnetization, can be strongly enhanced inside or in the vicinity of conducting nanoobjects, such as nanoparticles, nanodisks, and quantum dots, particularly near plasmon resonances~\cite{Magarill:1999,Tokman:1999,Hamidi:2015,Hurst:2018aa,Sinha-Roy:2020,Potashin:2020,Yang:2023,Kopasov:2026}. Experimentally, the photoinduced IFE magnetization is often probed by measuring the Faraday or Kerr rotation of a linearly polarized probe field. Indeed, recent pump--probe experiments have demonstrated sizable photoinduced Faraday rotation in conducting plasmonic systems~\cite{Cheng:2020aa,Han:2023,Gonzalez-Alcalde:2024,Parchenko:2025}. In particular, Han et al.~\cite{Han:2023}  reported remarkably large pump-induced Faraday rotation in graphene disk lattices under circularly polarized terahertz excitation, with rotation angles approaching 1$^\circ$ near the plasmon resonance. 

The experimental results motivate theoretical study of the pump-induced Faraday rotation in plasmonic systems, an effect which may be referred to as the orbital photoinduced Faraday effect. Although numerous studies have addressed theoretical calculations of the IFE magnetization itself, much less attention has been paid to the specific mechanisms responsible for the resulting Faraday rotation. An intuitive mechanism of Faraday rotation involves the Lorentz force exerted by the real magnetic field created by orbital currents. However, this magnetic field is known to be very small, of the order of 1--10~nT per 1~kW/cm$^2$ of the incident pump intensity even in  nanoparticles at the plasmon resonance~\cite{Hurst:2018aa,Potashin:2020,Cheng:2020aa,Gunyaga:2023}. This suggests that other possible mechanisms of Faraday rotation should be considered. 

\begin{figure}[ht]
	\centering
	\includegraphics[width=0.97\linewidth]{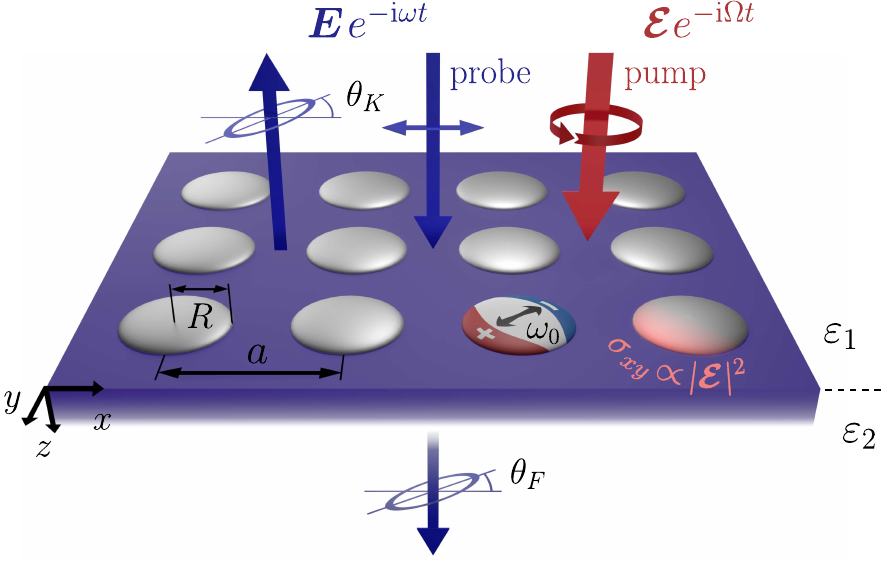}
	\caption{Photoinduced Faraday and Kerr effects in a square lattice of conducting disks. A circularly polarized pump wave 
	induces a high-frequency Hall conductivity of electrons in the disks, $\sigma_{xy} = -\sigma_{yx}$, resulting in Faraday and Kerr rotation of a linearly polarized probe wave.
	The rotation is resonantly enhanced when the pump and probe frequencies are close to the disk plasmon frequency $\omega_0$. 
}
\label{Fig1}
\end{figure}

Here, we study a different mechanism of the pump-induced Faraday rotation in conducting nanostructures. In this mechanism, the combined action of the circularly polarized pump field and the probe field directly induces an off-diagonal (Hall-like) component of the electron conductivity tensor at the probe frequency, $\sigma_{xy} = -\sigma_{yx}$~\cite{Durnev2023}.  Microscopically, the nonzero $\sigma_{xy}$ arises from  the third-order nonlinear response of the electron gas in the disks to the electric fields of the pump and probe beams. The resulting high-frequency Hall conductivity is proportional to the pump intensity: $\sigma_{xy} \propto |\bm{\mathcal E}|^2 P_{\rm circ}$, where $\bm{\mathcal E}$ is the amplitude of the pump electric field and $P_{\rm circ} = \pm 1$ corresponds to right- and left-handed circular polarization, respectively. In this sense, the pump field acts as an effective (synthetic) magnetic field leading to nonzero $\sigma_{xy}$, and, ultimately, to Faraday and Kerr rotation of the incident probe field. Importantly, however, the emerging conductivity $\sigma_{xy}$ is not caused by any actual magnetic field acting on electrons; instead, it arises from the third-order nonlinearity of the electron gas.

Based on this mechanism, we calculate the Faraday and Kerr response in a square lattice of conducting disks, Fig.~\ref{Fig1}. First, we derive analytical expressions for the Faraday and Kerr rotation angles and corresponding ellipticities for a general off-diagonal conductivity $\sigma_{xy}$. Then, we develop a microscopic theory of the pump-induced Hall conductivity $\sigma_{xy}$ for intraband electron transport which typically corresponds to microwave and terahertz ranges in semiconductors and up to optical and ultraviolet ranges in metals. The dominant contribution to $\sigma_{xy}$ originates from dynamic heating and cooling of the electron gas by the oscillating pump and probe fields and is governed by energy relaxation processes. 
We show that the Faraday and Kerr responses are substantially enhanced when the pump and probe frequencies are close to the plasmon resonance frequency of the disks. The enhancement factor compared with a uniform electron layer is $\sim Q^4$, where $Q$ is the quality factor of the plasmon resonance. We apply the developed theory to calculate the Faraday rotation at terahertz frequencies in a lattice of $n$-doped GaAs disks. The maximum calculated values of the Faraday angle are $\sim 0.1^\circ$ per 1~kW/cm$^2$ of the incident pump intensity. This is of the same order of magnitude as the experimentally observed values in graphene disk lattices in the same frequency range~\cite{Han:2023}. The effective magnetic field, which would produce the same rotation, is $\sim 0.1$~T, about six orders of magnitude larger than the actual magnetic field created by the IFE currents.

\section{Faraday rotation by a square lattice of conducting disks}

Consider a square lattice of conducting disks, with lattice period $a$, located in the plane $z = 0$ and surrounded by two half-spaces with dielectric constants $\eps_1$ at $z<0$ and $\eps_2$ at $z>0$, see Fig.~\ref{Fig1}. In what follows, we model the disks as thin oblate spheroids \cite{Allen:1983,Leavitt:1986,Mikhailov:1996} with in-plane radius $R$ and  out-of-plane semiaxis $h$, $h \ll R$. Within this model, a single disk screens the probe and pump electric fields such that the resulting electric field inside the disk remains uniform~\cite{landau8eng}. This allows one to find the plasmon frequency of the disk analytically; its value perfectly agrees with the numerical result for the abrupt-disk model~\cite{Fetter:1986}. In a lattice of disks, the electric field inside the disks becomes non-uniform, since in addition to external fields a given disk experiences electric fields of the charge redistributions induced in the other disks. However, in the limit $R/a \ll 1$ the latter fields are small and can be treated perturbatively, resulting in a plasmon-frequency shift proportional to $(R/a)^3$~\cite{Dahl:1992, Mikhailov:1996, Kosobukin:2016}. In the following, we consider the limit $R/a \ll 1$ and neglect the weak interdisk interaction.

The lattice is irradiated by normally incident probe and pump beams with frequencies $\omega$ and $\Omega$ and electric fields $\bm E^{\rm ext}(t) = \bm E^{\rm ext} e^{ - \i \omega t} + \mathrm{c.c}$ and $\bm{\mathcal E}^{\rm ext}(t) = \bm{\mathcal E}^{\rm ext} e^{- \i \Omega t} + \mathrm{c.c}$, respectively. The pump and probe fields induce electric currents in disks, which are periodic functions of the in-plane coordinate $\bm r = (x,y)$. The density of the current oscillating at the probe frequency, $\bm J(\bm r, z, t) = \bm J (\bm r, z) \e^{- \i \omega t} + \mathrm{c.c}$, is given by
\begin{equation}
\label{Jperiodic}
\bm J (\bm r, z) = \bm J \, \sum_{\bm A} \Theta(\bm r - \bm A, z)\:,
\end{equation} 
where the sum is taken over the lattice vectors $\bm A = a (n,m,0)$ with integer $n$ and $m$, and the function $\Theta(\bm r, z)$ equals unity if the point $(\bm r, z)$ lies inside the disk and is zero otherwise. The current density $\bm J$ inside a disk 
can be expressed in terms of the three-dimensional (3D) dynamic conductivity tensor $\sigma_{\alpha \beta}$ as:
\begin{equation}
\label{Jdisk}
J_{\alpha} = \sum_{\beta} \sigma_{\alpha \beta} E_\beta\:,
\end{equation}  
where $\alpha$ and $\beta$ run over the in-plane coordinates $x$ and $y$. Here, $\bm E$ is the uniform probe field inside the disks, which is equal to the sum of the incident field $\bm E^{\rm ext}$ and the field $\bm E^{\rm ind}$ induced by oscillating electric charges.

In the absence of the pump field, the conductivity tensor is diagonal, i.e. $\sigma_{xy} = \sigma_{yx} = 0$. A circularly polarized pump gives rise to the off-diagonal component $\sigma_{xy} = -\sigma_{yx}$, which is proportional to the pump intensity: $\sigma_{xy} \propto |\bm{\mathcal E}|^2 P_{\rm circ}$. Here, $\bm{\mathcal E}$ is the amplitude of the pump electric field inside the disks. 
The microscopic theory of $\sigma_{xy}$ is outlined in Sec.~\ref{sec:sigma_xy}.

The translation symmetry of the system in the $x$ and $y$ directions allows one to expand electric fields and currents in a Fourier series over reciprocal lattice vectors $\bm g = (2\pi/a) (n, m, 0)$. Here, we use the condition $h \ll R$, which allows one to treat the disks as quasi-two-dimensional objects located in the plane $z = 0$. The effective two-dimensional (2D) current density inside a disk is then given by:
\begin{equation}
\label{j2D}
\bm j(r) = \int \bm J\,\mathrm{d} z = 2 h \sqrt{1 - r^2/R^2}\, \bm J\:.
\end{equation}
Within this quasi-2D approximation, the Fourier series for the in-plane components of the electric field induced by currents and charges oscillating at the probe frequency, $\bm E^{\rm ind}(\bm r, z, t) = \bm E^{\rm ind}(\bm r, z) e^{-\i \omega t} + \mathrm{c.c.}$, reads:
\begin{equation}
\label{Fourier_E}
E_\alpha^{\rm ind}(\bm r, z) = \sum_{\bm g}  E_{\bm g, \alpha}^{\rm ind} \exp \bigl[ \i\bm g \cdot \bm r + \i q(z) |z|\bigr]\:.
\end{equation}
Here, $q(z)$ equals to $q_1$  for $z<0$ and $q_2$  for $z>0$, $q_{j} = \sqrt{\eps_j \omega^2/c^2 - g^2}$ ($j = 1,2$) and $E_{\bm g, \alpha}^{\rm ind}$ are the Fourier amplitudes of the induced electric field.

The amplitudes $E_{\bm g, \alpha}^{\rm ind}$ are proportional to the Fourier components $\bm j_{\bm g}$ of the effective 2D electric currents $\bm j(r)$. This relation is derived using the boundary conditions for the in-plane components of magnetic field at $z = 0$, which can be expressed through $E_{\bm g, \alpha}^{\rm ind}$ by solving the Maxwell equations at $z \neq 0$. It yields 
\begin{equation}
\label{Eg_jg}
E_{\bm g, \alpha}^{\rm ind} = \frac{4 \pi}{\omega} \frac{g_\alpha (\bm g \cdot \bm j_{\bm g})}{\eps_1 q_2 + \eps_2 q_1} - \frac{4\pi \omega}{c^2} \frac{j_{\bm g, \alpha}}{q_1 + q_2}\:,
\end{equation}
where 
\begin{equation}
\label{jg}
\bm j_{\bm g} = F \aver{ \bm j(r) \e^{-\i \bm g \cdot \bm r}}\:,
\end{equation}
$F = \pi R^2/a^2$ is the disk filling factor, and $\aver{\dots}$ denotes averaging over the in-plane area of a disk.

In the terahertz pump-probe experiments, the wavelength $\lambda$ of the incident fields is typically much larger than the lattice constant $a$. For example, in Ref.~\cite{Han:2023} $a = 1.5~\mu$m and $\lambda \approx 85~\mu$m (corresponding to the frequency 3.5~THz and $\eps_1 = 1$). As a result, $|\bm g| \gg \sqrt{\eps_j} \omega/c$ and, consequently, $q_j$ are imaginary for all $\bm g$ except $\bm g = 0$. Hence, in the far field zone, $E_\alpha^{\rm ind}(\bm r, z)$ has the form of outgoing plane waves with the wave vectors parallel to the $z$-axis. Setting $\bm g = 0$ in Eqs.~\eqref{Fourier_E} and \eqref{Eg_jg}, one obtains
\begin{equation}
\label{far_field}
\bm E^{\rm ind} (\mathrm{far~field}) = 
 - \frac{2\pi}{c \overline{n}} \,\bm j_{\bm g = 0} \exp\left[\i n(z) \omega |z|/c \right]\:,
 \end{equation}
 where $n_j = \sqrt{\eps_j}$ are the refraction indices of the surrounding media, $\overline{n} = (n_1+n_2)/2$, and $n(z)$ equals to $n_1$ for $z<0$ and $n_2$  for $z>0$. 
 
The Faraday and Kerr rotation originates from the different transmission and reflection of the right-hand and left-hand circularly polarized components of the probe field. We will further consider the low-intensity regime, when the pump-induced off-diagonal conductivity is small, $|\sigma_{xy}| \ll |\sigma_{xx}|$, and retain only linear in $\sigma_{xy}$ terms. In this case, the rotation angles and ellipticities of the transmitted and reflected probe waves are given by~\cite{Palik1970, OConnell1982, Glazov:2012aa}
\begin{equation}
\label{Faraday_Kerr}
\epsilon_F -\i \theta_F \approx \frac{t_+ - t_-}{t_+ + t_-}\:,~\epsilon_K -\i \theta_K \approx \frac{r_+ - r_-}{r_+ + r_-}\:.
\end{equation}
Here, $\theta_{F/K}$ and $\epsilon_{F/K}$ are the Faraday/Kerr rotation angles and the accompanying ellipticities, respectively, and $t_\pm$ and $r_\pm$ are the amplitude transmission and reflection coefficients of the circularly polarized components $E_{\rm circ^\pm} = (E_x \mp \i E_y)/\sqrt{2}$ of the probe field. To calculate $t_\pm$ and $r_\pm$, according to Eq.~\eqref{far_field}, one has to find the zeroth spatial Fourier harmonic of the current components $j_{\rm circ^\pm} = j_x \mp \i j_y$. Equations~\eqref{jg}, \eqref{j2D} and \eqref{Jdisk} yield
 \begin{equation}
 \label{jg0}
\left[ j_{\rm circ^\pm} \right]_{\bm g = 0} =  F \tilde{h} \,\sigma_\pm E_{\rm circ^\pm}\:,
 \end{equation}
 where $\tilde h = 4 h/3$ and $\sigma_\pm = \sigma_{xx} \pm \i \sigma_{xy}$. 

The quantities $E_{\rm circ^\pm}$ in Eq.~\eqref{jg0} are the circularly polarized components of the screened probe field inside a disk. These components are given by $E_{\rm circ^\pm} =  t_{12} \, E_{{\rm circ}^\pm}^{\rm ext} + E_{\rm circ^\pm}^{\rm ind}$, where $t_{12}=2n_1/(n_1+n_2)$ is the amplitude transmission coefficient for the light incident on the interface of two dielectric media in the absence of disks. The relation between $E_{\rm circ^\pm}^{\mathrm{ind}}$ and the components of the external field  $E_{\rm circ^\pm}^{\mathrm{ext}}$ is found from Eq.~\eqref{Fourier_E} taken at $z=0$ by multiplying both sides by $\nu(r) = 3/2\sqrt{1 - r^2/R^2}$ and averaging over the in-plane area of a disk~\cite{Mikhailov:1996}. As a result, one obtains the following relation:
\begin{equation}
\label{aver_sigma_E}
E_{\rm circ^\pm} =  t_{12} \, E_{{\rm circ}^\pm}^{\rm ext}/\zeta_\pm\:,
\end{equation}
where the response functions are~\cite{Mikhailov:1996}
\begin{equation}
\label{zeta}
\zeta_\pm = 1 + \frac{2\pi F \tilde{h}\,\sigma_\pm}{c \overline{n}} + \frac{\pi \i F \tilde{h} \, \sigma_\pm}{\omega \overline{\eps}} \sum_{\bm g \neq 0} g \left|\aver{\nu (r) \e^{\i \bm g \cdot \bm r}}\right|^2\:,
\end{equation}
and $\overline{\eps} = (\eps_1 + \eps_2)/2$. The second term in Eq.~\eqref{zeta} corresponds to $\bm g = 0$ in Eq.~\eqref{Fourier_E} and is responsible for the radiative decay of plasmons in the disk lattice. The third term corresponds to imaginary $q_1$ and $q_2$ in Eq.~\eqref{Fourier_E}, and hence describes the screening of the external field by evanescent fields induced by oscillating charges in the disks. It is this term that gives rise to the plasmon resonance. 

%
%
%


Substituting Eqs.~\eqref{jg0} and \eqref{aver_sigma_E} into Eq.~\eqref{far_field}, we obtain the coefficients   describing transmission and reflection of the probe field by the disk lattice:
\begin{equation}
\label{t_pm}
t_\pm = t_{12} \left( 1 - \frac{2\pi F \tilde{h}\,\sigma_\pm}{c \overline{n} \zeta_\pm} \right)\:,~~r_\pm = t_\pm - 1\:.
\end{equation}
To proceed further, we recall that the pump-induced off-diagonal conductivity is small, $|\sigma_{xy}| \ll |\sigma_{xx}|$, and we calculate the Faraday and Kerr rotation, Eq.~\eqref{Faraday_Kerr}, to the first order in $\sigma_{xy}$. In this limit, the diagonal component $\sigma_{xx}$ can be replaced by its unperturbed value in the Drude model, $\sigma_{xx} = \i n_{e} e^2/[m^*(\omega + \i \gamma)]$, where $n_{e}$ is the equilibrium (3D) density of electrons, $m^*$ is the effective electron mass, and $\gamma$ is the damping frequency. Substituting Eq.~\eqref{t_pm} into Eq.~\eqref{Faraday_Kerr}, using Eq.~\eqref{zeta} for $\zeta_\pm$, and evaluating the sum $\sum_{\bm g \neq 0} g \left|\aver{\nu (r) \e^{\i \bm g \cdot \bm r}}\right|^2 \approx 3 a^2/(4R^3)$ in the limit $R/a \ll 1$, we finally obtain
 \begin{equation}
\label{Faraday_final}
\epsilon_F -\i \theta_F = \frac{\Gamma \omega^2 (\omega + \i \gamma)\,\sigma_{xy}/\sigma_{xx}}{[\omega^2 - \omega_0^2 + \i \omega (\gamma + \Gamma)] (\omega^2 - \omega_0^2 + \i \omega \gamma)}\:,
\end{equation}
and 
\begin{equation}
\label{Kerr_final}
\epsilon_K -\i \theta_K = \frac{t_{12} (\omega^2 - \omega_0^2 + \i \omega \gamma)}{r_{12} (\omega^2 - \omega_0^2 + \i \omega \gamma) - \i \omega \Gamma}  (\epsilon_F -\i \theta_F) \:,
\end{equation}
where
\begin{equation}
\label{Gamma}
\Gamma = \frac{2 \pi F n_{\rm 2D} e^2}{m^* c\,\overline{n}}
\end{equation}
is the radiative decay rate of the disk lattice~\cite{Mikhailov:1996},
\begin{equation}
\label{omega0}
\omega_0^2 = \frac{3 \pi^2 n_{\rm 2D} e^2}{4 m^* \overline{\eps} R} \
\end{equation}
is the squared frequency of the dipole plasmon mode, $n_{\rm 2D} = n_{e} \tilde h$ is the effective 2D electron density, and $r_{12} = t_{12} - 1$.

Equations~\eqref{Faraday_final} and \eqref{Kerr_final} are the key result of this section. They give the Faraday and Kerr rotation angles and ellipticities of the incident probe field induced by the lattice of conducting disks. Notably, the rotation angles exhibit resonances at the plasmon frequency $\omega_0$. The resonances are related to the plasmon-induced enhancement of the probe electric field in the disks, see also Ref.~\cite{Sepulveda:2010}. For example, if the incident probe field is parallel to the $x$-axis, the enhanced field inside the disks is given by
\begin{eqnarray}
\label{Eprobe_plasmon}
E_x &=& \frac{\omega(\omega+\i\gamma)}{\omega^2 - \omega_0^2 + \i \omega (\gamma + \Gamma)} t_{12} E_x^{\rm ext}\:, \nonumber \\
E_y &=& \frac{\sigma_{xy}}{\sigma_{xx}} \frac{\omega(\omega+\i\gamma)(\i \Gamma \omega - \omega_0^2) }{[\omega^2 - \omega_0^2 + \i \omega (\gamma + \Gamma)]^2} t_{12} E_x^{\rm ext} \:.
\end{eqnarray}


		
The Faraday and Kerr rotations in Eqs.~\eqref{Faraday_final} and \eqref{Kerr_final} are governed by the pump-induced Hall conductivity $\sigma_{xy}$. The microscopic theory of $\sigma_{xy}$ is presented in the next section.
	
\section{Pump-induced Hall conductivity} \label{sec:sigma_xy}
	
Here, we develop a kinetic theory of the Hall conductivity $\sigma_{xy}$ induced by the circularly polarized pump field. We consider the classical regime when the photon energies $\hbar \omega$ and $\hbar \Omega$ are much smaller than the Fermi energy of electrons, which is relevant for terahertz spectral range in semiconductors and up to optical and ultraviolet ranges in metals. We treat electrons in the thin oblate spheroids as three-dimensional and consider the case in which both semiaxes of the spheroid, $R$ and $h$, are much larger than the mean free path of electrons $l_e$. This allows us to neglect the surface effects and calculate $\sigma_{xy}$ for a uniform 3D electron gas in the disks. The theory closely follows that developed in Ref.~\cite{Durnev2023} for an unconfined 2D electron gas. Earlier, a similar kinetic theory with a simplified model of electron relaxation was used to discuss the optically induced Faraday effect in graphene~\cite{Glazov2014}.

The theory is based on the solution of the kinetic Boltzmann equation for the electron distribution function $f(\bm p, t)$:
\begin{equation}
\label{Boltzmann}
\pderiv{f}{t}  + e \left[ \bm E \e^{-\i\omega t} + \bm{\mathcal E} \e^{-\i\Omega t} + \mathrm{c.c.} \right] \cdot \pderiv{f}{\bm p} = I\{ f\}\:,
\end{equation}
where $\bm p$ is the electron momentum, and $I\{ f\}$ is the collision integral. We recall that $\bm E$ and $\bm{\mathcal E}$ are the uniform electric fields of the probe and pump inside the disk, respectively. The relaxation of the angular harmonics of $f(\bm p, t)$ with the orbital  momenta $L =1$ and $L = 2$ is described by the energy-dependent times $\tau_1(\eps_{\bm p})$ and $\tau_2(\eps_{\bm p})$, given by $\tau_1^{-1} = \gamma = -\aver{\bm v I\{f\}}_{\bm p}/\aver{\bm v f}_{\bm p}$ and $\tau_2^{-1} = -\aver{v_x v_y I\{f\}}_{\bm p}/\aver{v_x v_y f}_{\bm p}$, respectively. Here, $\eps_{\bm p} = p^2/2m^*$ and $\bm v = \bm p/m^*$ are the electron energy and velocity, respectively, and $\aver{\dots}_{\bm p}$ denotes averaging over the directions of $\bm p$. The energy dissipation is given by $(\sum_{\bm p} \eps_{\bm p} f)/\tau_\eps$, where $\tau_\eps$ is the energy relaxation time.

Within the linear response, Eq.~\eqref{Boltzmann} at $\bm{\mathcal E} = 0$ yields electric current oscillating at frequency $\omega$ and determined by the diagonal Drude conductivity $\sigma_{xx} = \sigma_{yy}$. To account for modification of the conductivity tensor in the presence of the pump, one should go beyond the linear response and consider the third-order response, linear in the probe field and quadratic in the pump field. It gives rise to the transverse electric current, e.g., $J_y  \propto E_x |\bm{\mathcal E}|^2 P_{\rm circ}$, and, consequently, to the off-diagonal conductivity $\sigma_{yx} = -\sigma_{xy}$ proportional to the pump intensity. We therefore solve Eq.~\eqref{Boltzmann} perturbatively, by expanding the distribution function in the series in the electric field amplitudes $E$ and $\mathcal E$: 
\begin{multline} 
\label{perturbation_series}
f(\bm p, t) = f_0 (\eps_{\bm p}) + \left[f_{1\omega}(\bm p)\e^{-\i\omega t} + f_{1\Omega}(\bm p)\e^{-\i\Omega t} + \mathrm{c.c.}\right]  \\
+ \left[f_{2,\omega+\Omega}(\bm p)\e^{-\i(\omega+\Omega)t} + f_{2,\omega-\Omega}(\bm p)\e^{-\i(\omega-\Omega)t}+\mathrm{c.c.}\right]  \\
+ \left[f_{3,\omega}(\bm p)\e^{-\i\omega t}+\mathrm{c.c.}\right]\,.
\end{multline}
Here, $f_0$ is the equilibrium distribution function, $f_{1\omega} \propto E$ and $f_{1\Omega} \propto \mathcal E$ are the first-order corrections, which determine the Drude response, $f_{2,\omega+\Omega} \propto E \mathcal E$ and $f_{2,\omega-\Omega} \propto E \mathcal E^*$ are the second-order corrections and $f_{3,\omega} \propto E \mathcal E \mathcal E^*$ is the third-order correction, which determines the off-diagonal conductivity. 
	
We further calculate the transverse electric current $J_{y}$ induced by the $x$-component of the probe electric field $E_x$. The current is determined by the correction $f_{3,\omega}$ and reads
\begin{equation}
\label{jy0}
J_y = e \sum_{\bm p} v_y f_{3,\omega} (\bm p)\:.
\end{equation}	
The calculation of this current is presented in Appendix. The resulting expression for the Hall conductivity of  degenerate electron gas is
\begin{equation}
\label{sxy_general}
\frac{\sigma_{xy}}{\sigma_{xx}} = S(\omega,\Omega) - S(\omega,-\Omega)\:,
\end{equation}
where
\begin{multline} 
\label{S}
S(\omega,\Omega) = -\frac{\i e^2|\bm{\mathcal E}|^2P_{\mathrm{circ}} \left[2-\i(\omega-\Omega)\tau_1\right]}{6m^*(1+\i\Omega\tau_1)} \\
 \times \biggl[ (5 \tau_{1\omega}' + 2 \eps_F \tau_{1\omega}'') \tau_{\eps, \omega-\Omega} \\- 2\eps_F(\tau_{1\omega}' \tau_{2,\omega-\Omega})'
 - 5\tau_{1\omega}' \tau_{2,\omega-\Omega} \biggr]\:.
\end{multline}
Here, $\tau_{1\omega} = \tau_1/(1 - \i \omega \tau_1)$, $\tau_{n,\omega-\Omega} = \tau_n/[1 - \i (\omega-\Omega) \tau_n]$ ($n = 2,\eps$), $(\dots)' = d(\dots)/d\eps_{\bm p}$ denotes the derivative over electron energy, and the relaxation times and their energy derivatives are taken at the Fermi energy $\eps_F = \hbar^2(3 \pi^2 n_e)^{2/3}/2m^*$.

The contributions in the second and third lines of Eq.~\eqref{S} have different nature. The contribution in the second line, which is determined by the energy relaxation time $\tau_\eps$, is due to the dynamic heating and cooling of electrons by the oscillating pump and probe fields. The contributions in the third line are related to the optical alignment of electron momenta~\cite{Durnev2023}. At low temperatures, energy relaxation is much less efficient than momentum relaxation, so that $\tau_\eps \gg \tau_{1,2}$. Therefore, when the probe and pump frequencies are close, $\sigma_{xy}$ is dominated by the term in the second line of Eq.~\eqref{S}. This term is proportional to $\tau_{\eps, \omega-\Omega}$ and, hence, exhibits a resonance at $\omega \approx \Omega$ with the width $\tau_\eps^{-1}$. For $\tau_1 \propto (\eps_{\bm p})^{\alpha}$ where $\alpha$ is a real number, $\sigma_{xy}/\sigma_{xx}$ in the vicinity of the resonance is
\begin{multline}
\label{sxy_res}
	\frac{\sigma_{xy}}{\sigma_{xx}} \approx -\frac{\i e^2\tau_1|\bm{\mathcal E}|^2 P_{\mathrm{circ}} \, \alpha\bigl[(3+2\alpha) - \i\omega\tau_1(3-2\alpha)\bigr]}{3m^* \eps_F \bigl(1+\omega^2\tau_1^2\bigr)(1-\i\omega\tau_1)^2} \\ 
	\times \frac{\tau_\eps}{1-\i(\omega-\Omega)\tau_{\eps}}\:.
\end{multline}

Importantly, just as the probe field, the pump field $\bm{\mathcal E}$ inside the disks that drives the Hall response is also enhanced at the plasmon resonance: 
\begin{equation}
\label{Epump_res}
\bm{\mathcal E} = \frac{\Omega(\Omega+\i\gamma) }{\Omega^2 - \omega_0^2 + \i \Omega (\gamma + \Gamma)} t_{12} \bm{\mathcal E}^{\rm ext}\:,
\end{equation}
where we recall that  $\gamma = \tau_1^{-1}$.

\section{Results and discussion} \label{sec:results}

In this section, we apply Eqs.~\eqref{Faraday_final}, \eqref{Kerr_final}, \eqref{sxy_general} and \eqref{S} to calculate the pump-induced Faraday and Kerr rotation in the disk lattice. For calculations, we use parameters corresponding to $n$-doped GaAs disks on a substrate. Specifically, we take $m^* = 0.07\,m_0$, $\tau_1 = 0.2$~ps and $n_{e} = 3\times 10^{17}$~cm$^{-3}$, corresponding to the mobility $\mu \approx 5\times10^3$~cm$^2$/(V\,s), Fermi energy $\eps_F \approx 24\text{ meV}$ and $k_F l_e \approx 15$. We use the dielectric constants $\eps_1 = 1$ and $\eps_2 = 13$ and set the height of the disk $\tilde h = 0.1~\mu$m resulting in the effective 2D density $n_{\rm 2D} = 3\times 10^{12}$~cm$^{-2}$. The energy relaxation time is set to $\tau_\eps = 50\,\tau_1 = 10$~ps.

We perform calculations for lattices with three different disk radii $R =$~0.5, 1 and 2~$\mu$m. For each radius, the lattice period $a$ is chosen such that the ratio $R/a$ is fixed at 1/3 corresponding to the filling factor $F = \pi/9$. The radiative decay rate, Eq.~\eqref{Gamma}, is therefore the same for all three lattices, $\Gamma \approx 3.6\times10^{11}$~s$^{-1}$, so that $\Gamma/\gamma \approx 0.07$. The resulting plasmon frequencies, Eq.~\eqref{omega0}, for the chosen $R$ are $\omega_0/2\pi \approx 2.5$, 1.7 and 1.2 THz corresponding to $\omega_0/(\gamma + \Gamma) \approx 2.9$, 2 and 1.4. The intensity of the incident pump field, $\mathcal I = c n_1 |\bm{\mathcal E}^{\rm ext}|^2/2 \pi$ is set to $\mathcal I = 1$~kW/cm$^2$ corresponding to $|\bm{\mathcal E}^{\rm ext}| \approx 0.43$~kV/cm.

\subsection{Plasmon enhancement of the Faraday rotation}

There are two types of resonances in the spectra of the Faraday angle and ellipticity. The first one, at $\omega \approx \Omega$, originates from the off-diagonal conductivity $\sigma_{xy}$ of bulk electrons, see the second line of Eq.~\eqref{sxy_res}. This resonance is narrow because its width, $\tau_\eps^{-1}$, is determined by the slow energy relaxation processes. The second type of  resonance is due to excitation of plasmons in the disks. It occurs when either one or both of the frequencies  $\omega$ and $\Omega$ are close to the plasmon frequency $\omega_0$ and has a much larger width, $\gamma+ \Gamma$.  

\begin{figure}
	\centering
	\includegraphics[width=0.95\linewidth]{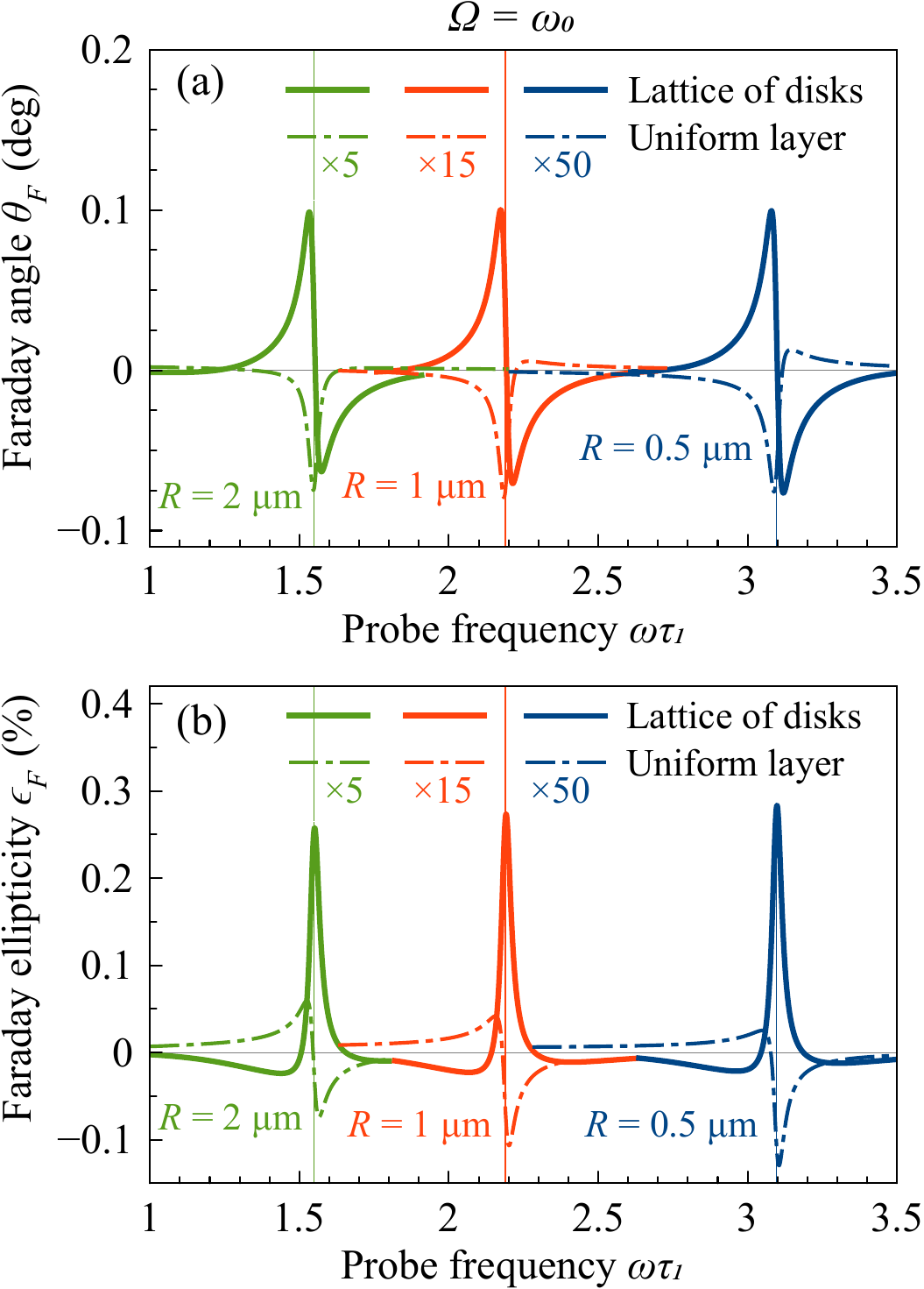}
	\caption{(a) Pump-induced Faraday rotation angle $\theta_F$ and (b) the accompanying ellipticity $\epsilon_F$ in   lattices of conducting disks for \textit{fixed pump frequencies} $\Omega$ set to be resonant with the lattice plasmon frequencies $\omega_0$ (shown by vertical lines). The three solid curves correspond to lattices with different disk radii $R =$~0.5, 1 and 2~$\mu$m, while the ratio $R/a$ is fixed at 1/3. For comparison, the dash-dotted lines show results for a uniform electron layer with the same average density of electrons per unit area as in the lattices. The dash-dotted lines are calculated for the same values of $\Omega$ as the corresponding solid lines.
	 The pump intensity is $\mathcal I = 1$~kW/cm$^2$, relaxation times $\tau_1 = \tau_2 \propto (\eps_{\bm p})^{-1/2}$; other parameters are listed in the text.
}
\label{Fig2}
\end{figure}

The narrow resonances are illustrated in Fig.~\ref{Fig2}. The results are calculated for $\tau_1 = \tau_2 \propto (\eps_{\bm p})^{-1/2}$ corresponding to electron scattering by short-range defects or acoustic phonons. The calculated Faraday angle and ellipticity are plotted 
as functions of the probe frequency $\omega$ for fixed pump frequencies $\Omega$ set to be resonant with the plasmon frequencies $\omega_0$ of the three lattices. It is seen that both the Faraday angle and ellipticity exhibit resonances at $\omega \approx \Omega$ with the width determined by $\tau_\eps^{-1}$. Since $\tau_\eps^{-1} \ll \gamma + \Gamma$, the resonances at $\omega \approx \Omega$ are obtained from Eqs.~\eqref{Faraday_final}, \eqref{sxy_res} and \eqref{Epump_res} with $\omega$ and $\Omega$ set equal to $\omega_0$ everywhere, except the second line of Eq.~\eqref{sxy_res}. It yields 
 \begin{equation}
\label{Faraday_res}
\epsilon_F -\i \theta_F \approx  \mathcal C(\omega_0) \frac{e^2 \gamma \Gamma\, t_{12}^2 |\bm{\mathcal E}^{\rm ext}|^2 P_{\mathrm{circ}}}{3 m^* \eps_F (\gamma+\Gamma)^3}  \frac{\tau_\eps}{1-\i(\omega-\Omega)\tau_{\eps}}\:,
\end{equation}
 where
 \begin{equation}
 \label{Cw}
 \mathcal C(\omega_0) =  - \alpha \frac{(3-2\alpha) \omega_0 + \i (3+2\alpha) \gamma}{(\omega_0 + \i \gamma)}\:.
 \end{equation}
The shape of the resonances in Fig.~\ref{Fig2} is determined by the phase of $\mathcal C(\omega_0)$. For $\alpha = -1/2$, relevant to short-range scatterers, the phase of $\mathcal C(\omega_0)$ is small for all $\omega_0$. As a result, the ellipticity $\epsilon_F$ is approximately proportional to the real part of the complex Lorentzian $\tau_\eps/[1-\i(\omega-\Omega)\tau_{\eps}]$, whereas the Faraday angle $\theta_F$ is proportional to its imaginary part. 

For comparison, the dash-dotted lines in Fig.~\ref{Fig2} show results for a uniform electron layer with the same average density of electrons per unit area as in the lattices, corresponding to the same $\Gamma$. The case of a uniform layer corresponds to the limit $\omega_0 \to 0$ in Eq.~\eqref{Faraday_final} and \eqref{Epump_res}. 
It is seen that both the Faraday rotation and ellipticity are substantially enhanced at the plasmon resonance in the disk lattice as compared to the uniform layer at the same frequency. The plasmon enhancement  originates from two factors. First, at the plasmon resonance, the electric currents that emit electromagnetic waves at the probe frequency are enhanced due to the increased electric field $\bm E$ inside the disks, Eq.~\eqref{Eprobe_plasmon}.
Second, the pump field $\bm{\mathcal E}$ itself is enhanced at the plasmon resonance, Eq.~\eqref{Epump_res}, which leads to an enhancement of the induced Hall conductivity $\sigma_{xy} \propto |\bm{\mathcal E}|^2$. In the limit $\omega_0 \gg \gamma + \Gamma$, the enhancement factor is 
\begin{equation}
\frac{(\epsilon_F - \i \theta_F)^{\rm lattice}}{(\epsilon_F - \i \theta_F)^{\rm layer}} \approx -\frac{\omega_0}{\gamma} \left( \frac{\omega_0}{\gamma + \Gamma} \right)^3\:,
\end{equation}
which scales approximately as $Q^4$, where $Q = \omega_0/(\gamma + \Gamma)$ is the quality factor of the plasmon resonance.

\begin{figure}
	\centering
	\includegraphics[width=0.95\linewidth]{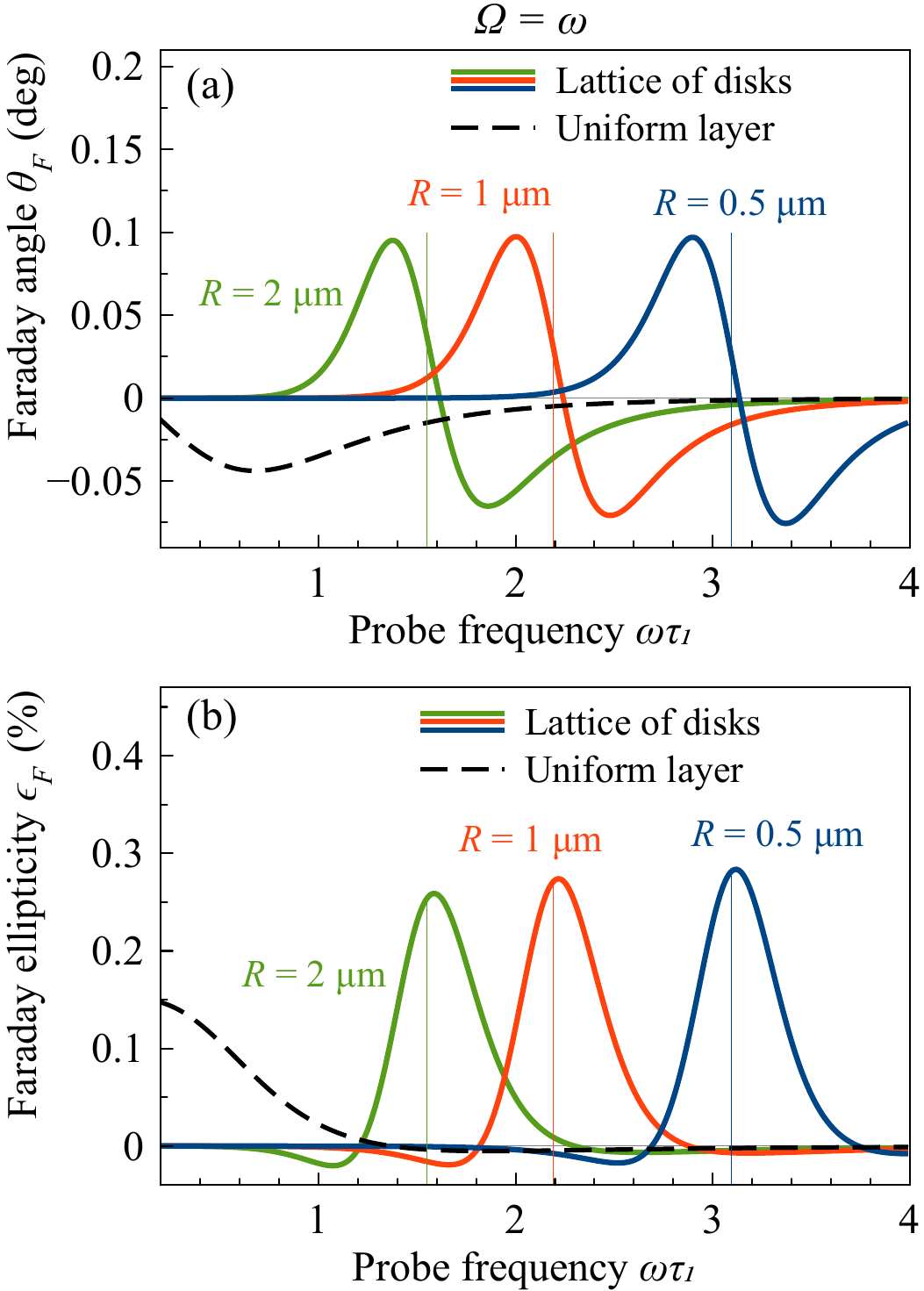}
	\caption{(a) Pump-induced Faraday rotation angle $\theta_F$ and (b) the accompanying ellipticity $\epsilon_F$ in   lattices of conducting disks \textit{for coinciding pump and probe frequencies}, $\omega = \Omega$.
The three solid curves correspond to lattices with different disk radii $R =$~0.5, 1 and 2~$\mu$m, while the ratio $R/a$ is fixed at 1/3. The lattice plasmon frequencies $\omega_0$ are shown by vertical lines.
For comparison, the dashed lines show the results for a uniform electron layer with the same average density of electrons per unit area as in the lattices. The pump intensity is $\mathcal I = 1$~kW/cm$^2$, relaxation times $\tau_1 = \tau_2 \propto (\eps_{\bm p})^{-1/2}$; other parameters are listed in the text.
}
\label{Fig3}
\end{figure}

The spectral dependences of the Faraday angle and ellipticity for coinciding pump and probe frequencies, $\Omega = \omega$, are plotted in Fig.~\ref{Fig3}. Both quantities indeed exhibit plasmon resonances with a larger, as compared to Fig.~\ref{Fig2}, width determined by the plasmon decay rate $\gamma + \Gamma$. For comparison, the results for a uniform electron layer are shown by the dashed lines. In fact, the maximum values of $\theta_F$ and $\eps_F$ achieved in the uniform electron layer are comparable to those in the disk lattices. However, these values are achieved only at near-zero or exactly zero frequencies. Thus, the disk lattice shifts the maximum response toward higher frequencies.

The Faraday angle and ellipticity are proportional to the radiative decay $\Gamma$, which is, in turn, proportional to the disk filling factor $F$, Eq.~\eqref{Gamma}. The results in Figs.~\ref{Fig2} and \ref{Fig3} are shown for the lattices with the identical filling factor, therefore the resonance amplitudes are the same for all three lattices. If instead the lattice period $a$ is fixed, the amplitudes decrease with decreasing disk radius $R$, since $F \propto R^2$.

\begin{figure}
	\centering
	\includegraphics[width=0.95\linewidth]{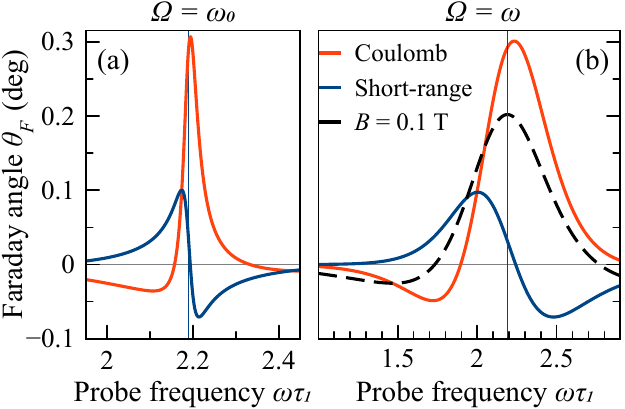}
	\caption{Role of the scattering potential. Pump-induced Faraday rotation angle (a) for a fixed pump frequency $\Omega = \omega_0$ and (b) for coinciding pump and probe frequencies, $\omega=\Omega$, for two types of scattering potential: short-range scatterers (blue solid curves) and charged scatterers with the Coulomb potential (red solid curves). For comparison, the black dashed line shows Faraday rotation by the same disk lattice in an external magnetic field $B = 0.1\text{ T}$. $R = 1$~$\mu$m, $\mathcal I = 1$~kW/cm$^2$.
	}
\label{Fig4}
\end{figure}

The results in Figs.~\ref{Fig2} and \ref{Fig3} are calculated for electron scattering by short-range defects or acoustic phonons resulting in $\tau_1 = \tau_2 \propto (\eps_{\bm p})^{\alpha}$ with $\alpha = -1/2$. However, the shapes and amplitudes of the resonances, determined by the coefficient $\mathcal C(\omega_0)$, Eq.~\eqref{Cw}, might be different for other scattering mechanisms. For comparison, in Fig.~\ref{Fig4} we present the calculated $\theta_F$ for electron scattering by charged impurities with the Coulomb potential, which corresponds to $\tau_1 = 3\tau_2 \propto (\eps_{\bm p})^{3/2}$ and hence $\alpha = 3/2$. It is seen that both the shape and the amplitude of the resonances depend on the scattering mechanism. In particular, for Coulomb scatterers, Fig.~\ref{Fig4}(b) shows that the maximum Faraday rotation is achieved when the pump and probe frequencies are almost equal to the plasmon frequency $\omega_0$. In contrast, for short-range scatterers the two maxima are achieved at frequencies detuned from the plasmon frequency.


The maximum values of the Faraday angle in Figs.~\ref{Fig2}--\ref{Fig4} are of the order of $0.1^\circ$ per 1~kW/cm$^2$ of the terahertz  pump intensity. Although the presented theory is developed for 3D electrons in thin oblate spheroids, we expect $\theta_F$ of the same order of magnitude for a lattice of strictly 2D graphene disks with comparable electron sheet density, geometric parameters and plasmon quality factor, studied experimentally in Ref.~\cite{Han:2023}. Taking into account the pump-pulse intensity  $\mathcal I \sim 10$~kW/cm$^2$ used in Ref.~\cite{Han:2023}, we estimate $\theta_F \sim 1^\circ$, in agreement with experimental observations.

So far, we have discussed the Faraday angle and ellipticity. The Kerr rotation angle and ellipticity are given by Eq.~\eqref{Kerr_final}. For the chosen parameters, the factor that determines the ratio between the Kerr and Faraday responses is almost frequency independent, resulting in $\epsilon_K -\i \theta_K \approx (t_{12}/r_{12}) (\epsilon_F -\i \theta_F)$.

\subsection{Effective vs. real magnetic field}

The calculated pump-induced Faraday and Kerr responses can be compared with those produced by an external magnetic field $\bm B \parallel z$. The latter can be found from the general Eqs.~\eqref{Faraday_final} and \eqref{Kerr_final} with $\sigma_{xy}/\sigma_{xx} = \omega_c \tau_{1\omega}$, where $\omega_c = e B/(m^* c)$ is the cyclotron frequency.
The resulting dependence of the Faraday angle for $B = 0.1$~T is shown by the dashed curve in Fig.~\ref{Fig4}(b). It is seen that the calculated pump-induced Faraday rotation for $\mathcal I = 1$~kW/cm$^2$ is similar to that produced by a magnetic field $B \sim 0.1$~T. The external magnetic field that produces the same Faraday rotation as the pump field can be estimated by equating $\sigma_{xy}/\sigma_{xx}$ given by Eq.~\eqref{sxy_res} to $\omega_c \tau_{1\omega}$. At $\omega \approx \Omega$ and $\omega \tau_1 \gg 1$, such an effective (synthetic) magnetic field is estimated as
\begin{equation}
\label{Bsyn}
B_{\rm syn} \sim \frac{ec\,\tau_\eps |\bm{\mathcal E}|^2 }{\eps_F (\omega \tau_1)^\beta} P_{\rm circ}\:,
\end{equation}
where $\beta = 2$ for short-range scatterers and $\beta = 3$ for Coulomb scatterers.

For comparison, let us estimate the actual magnetic field produced by the vortex currents in the IFE. The circularly polarized pump field induces a dc edge current circulating along the disk perimeter, $J_{\rm IFE} \propto |\bm{\mathcal E}|^2 P_{\rm circ}$, which, in the limit $\omega\tau_1 \gg 1$, is given by~\cite{Durnev2021,Gunyaga:2023}
\begin{equation}
J_{\rm IFE} = \frac{e^3 n_{\rm 2D} }{(m^*)^2 \omega^3} |\bm{\mathcal E}|^2 P_{\rm circ}\;.
\end{equation}
This current produces the magnetic field at the center of the disk, $B_{\rm IFE} = 2 \pi J_{\rm IFE}/(c R)$. The ratio of this field to the synthetic magnetic field, Eq.~\eqref{Bsyn}, can be estimated for Coulomb scatterers as
\begin{equation}
\label{Bratio}
 \frac{B_{\rm IFE}}{B_{\rm syn}} \sim \frac{2 \pi \sigma_{\rm 2D}}{c}\times \frac{l_e}{R} \times \frac{\tau_1}{\tau_\eps} \times \frac{v_F}{c}\:,
\end{equation}
where $v_F$ is the Fermi velocity, $l_e = v_F \tau_1$ is the mean free path of electrons, and $\sigma_{\rm 2D} = e^2 n_{\rm 2D} \tau_1/m^*$. Notably, all factors in Eq.~\eqref{Bratio} are either smaller than or much smaller than unity. For the used parameters, $2 \pi \sigma_{\rm 2D}/c \approx 0.5$, $l_e/R \approx 0.07$, $\tau_1/\tau_\eps = 0.02$ and $v_F/c \approx 10^{-3}$, resulting in $B_{\rm IFE}/B_{\rm syn} \sim 10^{-6}$.

\section{Conclusions}

To conclude, we have theoretically studied  pump-probe Faraday and Kerr effects in lattices of conducting disks resonantly enhanced by excitation of localized plasmons.
Treating the disks as thin oblate spheroids, we have derived analytical expressions for the Faraday and Kerr rotation angles and accompanying ellipticities in the metamaterial regime, when the pump and probe wavelengths exceed the lattice constant. The Faraday and Kerr responses are governed by the off-diagonal (Hall-like) conductivity $\sigma_{xy}$ of the electron gas, which is induced by the pump field. We have studied the kinetic mechanism of such a pump-induced Hall conductivity, in which $\sigma_{xy}$ originates from the third-order nonlinear response of the electron gas to the combined action of pump and probe electric fields. Analytical expressions for $\sigma_{xy}$ in a 3D electron gas are derived. The dominant contribution to $\sigma_{xy}$ originates from dynamic heating and cooling of the electron gas by high-frequency pump and probe electric fields. The Hall conductivity exhibits a narrow resonance when the pump and probe frequencies are close to each other; its width is determined by the energy relaxation rate.

The principal effect of the disk lattice is to shift the resonance of the Faraday and Kerr responses from the zero-frequency to the terahertz, infrared, or even optical frequency ranges, thereby enabling giant rotation angles that would otherwise require pump intensities orders of magnitude higher in uniform structures. When both the pump and probe frequencies are tuned close to the plasmon resonance, the rotation is enhanced by a factor of approximately $Q^4$, where $Q$ is the quality factor of the plasmon resonance. Both the shape and the amplitude of resonances in the spectra of the Faraday angle and ellipticity depend on the mechanism of electron scattering.

 The calculated Faraday rotation in a lattice of $n$-doped GaAs disks reaches $\sim 0.1^\circ$ per 1~kW/cm$^2$ of incident pump intensity, which is of the same order of magnitude as the values experimentally observed in graphene disk lattices in the terahertz range. The effective magnetic field corresponding to the same Faraday rotation angle is shown to be about six orders of magnitude larger than the actual magnetic field generated by the IFE currents circulating along the disk perimeter. 

Summing up, the proposed mechanism, in which Faraday rotation originates from the nonlinear pump-induced Hall conductivity $\sigma_{xy}$ rather than from the real magnetic field generated by circulating IFE currents, can be a major source of Faraday rotation in conducting nano- and microstructures.

\acknowledgements

We thank S. A. Tarasenko for fruitful discussions.
We acknowledge financial support from the Russian Science Foundation Grant No. 25-72-10031 and the Foundation for the Advancement of Theoretical Physics and Mathematics ``BASIS''.

\section*{Appendix: Calculation of $\sigma_{xy}$} 
\setcounter{equation}{0}
\renewcommand{\theequation}{A\arabic{equation}}

Here, we derive Eqs.~\eqref{sxy_general} and \eqref{S} of the main text for the pump-induced off-diagonal conductivity $\sigma_{xy}$.
Corrections to electron distribution function in Eq.~\eqref{perturbation_series} satisfy the following equations: 
\begin{subequations}
	\begin{align}
		 -\i\omega f_{1\omega} + e\bm E \pderiv{f_0}{\bm p} & = I\{f_{1\omega}\}\,, \label{app_eq_f1}\\
			-\i(\omega - \Omega)f_{2,\omega-\Omega} + e\!\left(\bm E\,\pderiv{f_{1\Omega}^*}{\bm p} + \bm{\mathcal E}^*\pderiv{f_{1\omega}}{\bm p}\right) &= I\{f_{2,\omega-\Omega}\},\label{app_eq_f2m} \\
					-\i(\omega + \Omega)f_{2,\omega+\Omega} + e\!\left(\bm E\,\pderiv{f_{1\Omega}}{\bm p} + \bm{\mathcal E}\,\pderiv{f_{1\omega}}{\bm p}\right) &= I\{f_{2,\omega+\Omega}\},\label{app_eq_f2p} \\
					-\i\omega f_{3,\omega} + e\left(\bm{\mathcal E} \pderiv{f_{2,\omega-\Omega}}{\bm p} + \bm {\mathcal E}^* \pderiv{f_{2,\omega+\Omega}}{\bm p} \right) &= I\{f_{3,\omega}\}\:. \label{app_eq_f3}
	\end{align}
	\label{app_equations_for_corrections}
\end{subequations} 
These equations are obtained by substituting Eq.~\eqref{perturbation_series} into Eq.~\eqref{Boltzmann} and treating the terms proportional to $\bm E$ and $\bm{\mathcal E}$ perturbatively. Equation for $f_{1\Omega}$ is obtained from Eq.~\eqref{app_eq_f1} by replacing $f_{1\omega}$ with $f_{1\Omega}$ and $\bm E$ with $\bm{\mathcal E}$ .
	
Averaging Eq.~\eqref{app_eq_f3} multiplied by $v_y$ over the directions of electron momentum $\bm p$, one obtains
\begin{multline}
\label{aver_vy_f3}
	\aver{v_y f_{3,\omega}}_{\bm p} = -e\tau_{1\omega} \aver{v_y \left(\bm{\mathcal E} \pderiv{f_{2\omega-\Omega}}{\bm p} + \bm{\mathcal E}^* \pderiv{f_{2\omega+\Omega}}{\bm p} \right)}_{\bm p}\:.
\end{multline}
The current density $J_y$ given by Eq.~\eqref{jy0} is then obtained by summing Eq.~\eqref{aver_vy_f3} over $\bm p$ and integrating by parts, which yields
\begin{equation}
	J_y = e^2\sum\limits_{\bm p}(f_{2,\omega-\Omega}\bm{\mathcal E} + f_{2,\omega+\Omega}\bm{\mathcal E}^*)\cdot\pderiv{(v_y\tau_{1\omega})}{\bm p}\,.
\end{equation}
Explicit calculation of the derivative in the right-hand side leads to the expression
\begin{multline}
	J_y = \frac{e^2}{3m^*}\sum\limits_{\bm p}(3\tau_{1\omega} + 2\eps_{\bm p}\tau_{1\omega}')\bigl(f_{2,\omega-\Omega}{\mathcal E}_y + f_{2,\omega+\Omega}{\mathcal E}^*_y\bigr)\\
	+ \frac{e^2}{3}\sum\limits_{\bm p}\tau_{1\omega}'\left[f_{2,\omega-\Omega}\left(3v_xv_y\mathcal E_x - \left(v_x^2 + v_z^2 - 2v_y^2\right)\!\mathcal E_y\right)\right.\\ \left. + f_{2,\omega+\Omega}\left(3v_xv_y\mathcal E^*_x - \left(v_x^2 + v_z^2 - 2v_y^2\right)\!\mathcal E^*_y\right)\right].
	\label{app_jy_via_f2}
\end{multline}

To proceed further, we calculate several sums involving corrections $f_{2,\omega\mp\Omega}$. Multiplying  Eq.~\eqref{app_eq_f2m} by $(3\tau_{1\omega} + 2\eps_{\bm p}\tau_{1\omega}')$, summing over $\bm p$ and integrating by parts, one obtains
\begin{multline}
	\sum\limits_{\bm p}(3\tau_{1\omega} + 2\eps_{\bm p}\tau_{1\omega}')f_{2,\omega-\Omega} = e\tau_{\eps,\omega-\Omega}(5\tau_{1\omega}' + 2\eps_{F}\tau_{1\omega}'')\\ \times\sum\limits_{\bm p}\bigl[(\bm v\cdot\bm E)f^*_{1\Omega} + (\bm v\cdot\bm{\mathcal E}^*)f_{1\omega}\bigr]\:.
	\label{app_sum_0th_harmonic}
\end{multline}
Multiplying Eq.~\eqref{app_eq_f2m} by either $(v_x^2 + v_z^2 - 2v_y^2)$ or $v_xv_y$ and averaging over the direction of $\bm p$, we obtain
\begin{multline}
\left<v_xv_yf_{2,\omega-\Omega}\right>_{\bm p} \\ = -e \tau_{2,\omega-\Omega} \aver{v_x v_y \left(\bm E\,\pderiv{f_{1\Omega}^*}{\bm p} + \bm{\mathcal E}^*\pderiv{f_{1\omega}}{\bm p}\right)}_{\bm p}\:,
\end{multline}
and analogous expression for $\left<(v_x^2 + v_z^2 - 2v_y^2) f_{2,\omega-\Omega}\right>_{\bm p}$.
Multiplying these averages by $\tau_{1\omega}'$, summing over $\bm p$ and integrating by parts, one obtains
\begin{subequations}
	\begin{align}
			&\begin{multlined}
			\sum\limits_{\bm p} \tau_{1\omega}'(v_x^2 + v_z^2 - 2v_y^2)f_{2,\omega-\Omega} = e\sum\limits_{\bm p}\left(\bm E f^*_{1\Omega} + \bm{\mathcal E}^*f_{1\omega}\right)\\[-2ex]
			 \cdot\!\pderiv{}{\bm p}\left[\tau_{1\omega}'\tau_{2,\omega-\Omega}\left(v_x^2 + v_z^2 - 2v_y^2\right)\right]\:,\hskip3ex
		\end{multlined}\\[1ex]	
	&\begin{multlined}
			\sum\limits_{\bm p} \tau_{1\omega}'v_xv_yf_{2,\omega-\Omega} \\[-2ex]= e\sum\limits_{\bm p}\left(\bm E f^*_{1\Omega} + \bm{\mathcal E}^*f_{1\omega}\right)\cdot\pderiv{}{\bm p}\bigl[\tau_{1\omega}'\tau_{2,\omega-\Omega}\,v_xv_y\bigr]\:.
		\end{multlined}
	\end{align}
\end{subequations}
Calculating the derivatives in the right-hand sides and using the relation (and analogously for $f_{1\Omega}$)
\begin{equation}
\label{aver_v3_f1}
	\left<v_{\alpha}v_{\beta}v_{\gamma} f_{1\omega}\right>_{\bm p} = \frac{v^2}{5}\bigl<(v_{\alpha}\delta_{\beta\gamma} + v_{\beta}\delta_{\alpha\gamma} + v_{\gamma}\delta_{\alpha\beta})f_{1\omega}\bigr>_{\bm p}\:,
\end{equation}
we obtain
\begin{subequations}
	\begin{align}
		&\begin{multlined}
			\!\!\!\!\sum\limits_{\bm p} \tau_{1\omega}'\bigl(v_x^2 + v_z^2 - 2v_y^2\bigr)f_{2,\omega-\Omega}  \\[-1.5ex]=
			\frac{2e}{5m^*}\bigl[5(\tau_{1\omega}'\tau_{2,\omega-\Omega}) + 2\eps_{F}(\tau_{1\omega}'\tau_{2,\omega-\Omega})'\bigr]\\
			\times\sum\limits_{\bm p}\left[v_xE_x f^*_{1\Omega} + (v_x\mathcal E^*_x - 2v_y\mathcal E^*_y)f_{1\omega}\right],\!\!\!\!
		\end{multlined}\\[1ex]
		&\begin{multlined}
			\!\!\!\!\sum\limits_{\bm p} \tau_{1\omega}'v_xv_yf_{2,\omega-\Omega} \\[-1.5ex]= \frac{e}{5m^*}\bigl[5(\tau_{1\omega}'\tau_{2,\omega-\Omega}) + 2\eps_{F}(\tau_{1\omega}'\tau_{2,\omega-\Omega})'\bigr] \\
			\times\sum\limits_{\bm p}\left[v_yE_xf^*_{1\Omega} + (v_y\mathcal E^*_x + v_x\mathcal E^*_y)f_{1\omega}\right].\!\!\!\!
		\end{multlined}
	\end{align}
	\label{app_sums_2nd_harmonic}
\end{subequations}
Relation~\eqref{aver_v3_f1} follows from the fact that the first-order corrections $f_{1\omega} \propto \bm v \cdot \bm E$ and $f_{1\Omega} \propto \bm v \cdot \bm{\mathcal E}$ and, hence, contain only the first angular harmonic in  momentum space. Expressions in the second lines of Eq.~\eqref{app_sums_2nd_harmonic} are taken at the Fermi energy. Analogous sums involving $f_{2,\omega+\Omega}$ are obtained from Eqs.~\eqref{app_sum_0th_harmonic} and \eqref{app_sums_2nd_harmonic} by replacing $-\Omega$ with $\Omega$, $\bm{\mathcal E}^*$ with $\bm{\mathcal E}$ and $f_{1\Omega}^*$ with $f_{1\Omega}$.

Finally, taking into account that the following sums give the first-order currents linear in electric fields:
\begin{equation}
	e\sum\limits_{\bm p}\bm v f_{1\omega} = \sigma_{xx}\bm E\,,\qquad e\sum\limits_{\bm p}\bm v f_{1\Omega} = \sigma_{xx} \frac{\tau_{1\Omega}}{\tau_{1\omega}}\bm{\mathcal E}\,,
\end{equation}
substituting Eqs.~\eqref{app_sum_0th_harmonic} and \eqref{app_sums_2nd_harmonic} into Eq.~\eqref{app_jy_via_f2} for the current and using $\bm{\mathcal E} = | \bm{\mathcal E}| (1,\i P_{\mathrm{circ}}, 0)/\sqrt{2}$, we calculate the transverse Hall current $J_y$ and therefore $\sigma_{xy}$ given by Eqs.~\eqref{sxy_general} and \eqref{S} of the main text.

\bibliographystyle{apsrev4-1-customized}
\bibliography{bibliography}

\end{document}